\documentclass{article}
\usepackage{epsfig}

\begin{document}

\title{\Large \bf An ``all--poles'' approximation to collinear resummations in the Regge limit of perturbative QCD}
\author{{Agust{\'\i}n~Sabio~Vera\thanks{Alexander von Humboldt Research Fellow}}\\[2.5ex] {\it II. Institut f{\"u}r Theoretische Physik, Universit{\"a}t Hamburg}\\{\it Luruper Chaussee 149, 22761 Hamburg, Germany}}

\maketitle

\vspace{-8cm}
\begin{flushright}
{\small DESY--05--xxx}
\end{flushright}

\vspace{7cm}
\begin{abstract} 
The procedure to improve the convergence in transverse momentum space of the 
NLL BFKL kernel using a $\omega$--shift is revisited. An accurate 
approximation to this shift only depending on transverse momenta is presented. 
This approximation is based on a Bessel function of the first kind with 
argument depending on the strong coupling and a double logarithm of the 
ratio of transverse scales. A comparison between different renormalization 
schemes is also included. 
\end{abstract}

\section{Introduction}
In recent years there has been an intense activity trying to understand which  
are the effective degrees of freedom driving the strong interaction at large 
energies. For processes where the center--of--mass energy is much larger than 
the other scales in the scattering an instructive picture emerges from the 
Balitsky--Fadin--Kuraev--Lipatov (BFKL) approach~\cite{FKL}. The effective 
picture in this framework relies upon $t$--channel ``Reggeized'' gluons 
interacting with each other via standard gluons in the $s$--channel. Although 
this simple structure must be modified at higher energies in order 
to introduce unitarization corrections, there 
is a window at present and future colliders where the BFKL predictions should 
hold.

The leading logarithmic (LL) approximation in BFKL resums terms of the form 
$(\alpha_s \ln{s})^n$ to all orders. In such a limit diagrams contributing to 
the running of the strong coupling do not appear and $\alpha_s$ remains as a 
constant free parameter. Also the factor needed to scale the 
energy in the logarithms is not fixed, therefore the predictability of 
the LL approximation is limited. The situation improves already in the  
next--to--leading logarithmic (NLL) approximation 
where diagrams with an extra power in the
 coupling without introducing an extra logarithm in energy are taken into 
account. In this case the coupling is allowed to run and the energy scale 
is fixed.

The calculation of the NLL corrections to the BFKL equation was 
lengthy and completed in Ref.~\cite{FLCC}. It was needed to evaluate  
corrections to the gluon trajectory~\cite{trajectory} and real emission 
kernel~\cite{emission}, as well 
as to check that the bootstrap conditions are fulfilled in order to ensure 
that the complete set of NLL contributions are indeed correctly 
resummed~\cite{Bootstrap}.

Lately, to be able to perform phenomenological studies, the 
NLL calculation and numerical implementation of 
particular impact factors, describing the nature of the interaction between 
the Reggeized gluons and the scattering particles for different physical 
processes, is a remaining challenge~\cite{impactfactors}.

In the construction of BFKL cross--sections the gluon Green's function is the 
element carrying the energy dependence and therefore it deserves a careful 
study. There has been a lot of theoretical work trying to understand its 
structure which, since the appearance of the NLL kernel, has revealed as quite 
non--trivial. It was shown in Ref~\cite{Ross} that, in the transverse momentum 
space, the NLL Green's function presents some instability. The origin of 
this instability and more convergent kernels including renormalization--group (RG) generated terms 
were presented in Ref.~\cite{Salam}. Based on this type of RG--improved 
kernels, including more sophisticated approaches, several studies of the 
Green's function have been 
performed in Ref.~\cite{Ciafalonietal}. Other analysis of different aspects 
of the NLL kernel can be found in Ref.~\cite{NLOpapers}.

The aim of the present work is to revisit the approach of Ref.~\cite{Salam} and to 
extract the structure in transverse momentum space of the double logarithms there resummed. 
In order to do so, an approximation to the original $\omega$--shift  
will be performed. It will be shown how this approximation is an accurate one and the 
corresponding expression in transverse momentum space very simple. The main feature of the 
new RG--improved kernel will be that it does not mix transverse with longitudinal momentum 
components. From the practical point of view it also allows for its straightforward implementation 
in the method of solution of the NLL BFKL equation proposed in Ref.~\cite{us} (for reviews 
see Ref.~\cite{reviews}).

In Section~\ref{NLOkernel} the NLL BFKL kernel will be presented and its representation 
in $\gamma$--space will be introduced. The particular RG--based scheme under investigation  
will be described at the end of this section. In Section~\ref{AllPoles} the approximated 
solution to this scheme will be derived and an analysis of its accuracy in the 
minimal substraction ($\overline{\rm MS}$) renormalization scheme will be shown. 
In Section~\ref{AporBessel} the corresponding RG--improved kernel is derived 
in transverse momentum space and its asymptotic behaviour in terms of the ratio of 
transverse scales discussed. Section~\ref{abitmore} is a brief one analysing the 
effect of changing the renormalization to the gluon--bremsstrahlung (GB) scheme. Finally, 
the Conclusions are presented.

\section{NLL BFKL kernel and modifications based on the renormalization 
group}
\label{NLOkernel}

The starting point of this analysis is the following representation of the 
NLL BFKL kernel acting on a smooth function~\cite{FLCC}:
\begin{eqnarray}
\label{ktKernel}
\int d^2 \vec{q}_2 \, {\cal K}\left(\vec{q}_1,\vec{q}_2\right) 
f\left({q}_2^2\right) = \\
&&\hspace{-5cm}\int \frac{d^2 \vec{q}_2}{\left|{q}_1^2-{q}_2^2\right|} \left\{\left[{\bar \alpha}_s+{\bar \alpha}_s^2\left({\cal S}-\frac{\beta_0}{4 N_c}\ln{\left(
\frac{\left|{q}_1^2-{q}_2^2\right|^2}
{{\rm max}\left({q}_1^2,{q}_2^2\right)\mu^2}\right)}
\right)\right]\right.\nonumber\\
&&\hspace{-5cm}\times\left.\left(f\left({q}_2^2\right)-2 \frac{{\rm min}\left({q}_1^2,{q}_2^2\right)}{\left({q}_1^2+{q}_2^2\right)}f\left({q}_1^2\right)\right)-\frac{{\bar \alpha}_s^2}{4}\left({\cal T}\left({q}_1^2,{q}_2^2\right)+\ln^2{\left(\frac{{q}_1^2}{{q}_2^2}\right)}\right)f\left({q}_2^2\right)\right\}, 
\nonumber
\end{eqnarray}
where $\bar{\alpha}_s \equiv g_\mu^2 N_c / (4 \pi^2)$, 
$\beta_0 = \left(11 N_c - 2 n_f\right)/3$,
${\cal S} = \left(4-\pi^2 +5 \beta_0/N_c \right)/12$,  and the function 
${\cal T} ({q}_1^2,{q}_2^2)$ can be obtained from 
Ref.~\cite{FLCC}. As it stands this expression is written in the 
${\overline{\rm MS}}$ renormalisation scheme. Alternatively, as in other 
calculations where a resummation of soft gluons is involved, the GB  scheme~\cite{GBesquema} can 
also be used:
\begin{eqnarray}
{\bar \alpha}_s^{\rm GB} &=& {\bar \alpha}_s \left(1+ {\cal S} \,{\bar \alpha}_s\right).
\end{eqnarray}
Both schemes will be under analysis in this work.

As it is well--known 
the collinear structure of this kernel can be extracted by projecting 
it on the modified LL eigenfunctions $\left({\bar \alpha}_s \left({q}^2\right)/{\bar \alpha}_s(\mu^2)\right)^{-1/2} {q}^{2(\gamma -1)}$, 
{\it i.e.}
\begin{eqnarray}
\label{nondiag}
\int d^2 \vec{q}_2 \, {\cal K} \left(\vec{q}_1,\vec{q}_2\right)
\left(\frac{{\bar \alpha}_s \left({q}_2^2\right)}{{\bar \alpha}_s \left({q}_1^2\right)}\right)^{-\frac{1}{2}}
\left(\frac{{q}_2^2}{{q}_1^2}\right)^{\gamma-1} = 
{\bar \alpha}_s \left({q}_1^2\right) \chi_0\left(\gamma\right)
+ {\bar \alpha}_s^2 \chi_1 \left(\gamma\right).
\end{eqnarray}
Note that the prefactor $\left({\bar \alpha}_s \left({q}_2^2\right)/{\bar \alpha}_s({q}_1^2)\right)^{-1/2} \simeq 1 + {\bar \alpha}_s \left({q}_1^2\right) \left(\beta_0 / 8 N_c\right) \ln{({q}_2^2/{q}_1^2)}$ generates a NLL term which makes the $\gamma$--representation of the kernel invariant under the 
$\gamma \rightarrow 1-\gamma$ transformation~\cite{FLCC}. 

Although ${\bar \alpha}_s \left({q}_1^2\right)
\simeq {\bar \alpha}_s \left(\mu^2\right) \left(1 - {\bar \alpha}_s \left(\mu^2\right) \frac{\beta_0}{4 N_c} \ln{\frac{{q}_1^2}{\mu^2}}\right)$ 
breaks the scale invariance, for the collinear analysis of the kernel it 
is enough to work with the scale invariant pieces which, at LL and 
NLL accuracy, read
\begin{eqnarray}
\label{LOeigen}
\chi_0 \left(\gamma\right) &=& 2 \psi(1) - \psi \left(\gamma\right)-
\psi \left(1-\gamma\right), 
\end{eqnarray}
\begin{eqnarray}
\label{chi1}
\chi_1 \left(\gamma\right) &=& {\cal S} \chi_0 \left(\gamma\right)
+ \frac{1}{4}\left(\psi''\left(\gamma\right)+ \psi''\left(1-\gamma\right)\right)
- \frac{1}{4}\left(\phi\left(\gamma\right)+ \phi\left(1-\gamma\right)\right)\\
&&\hspace{-2cm}-\frac{\pi^2 \cos{(\pi \gamma)}}{4 \sin^2(\pi \gamma)(1-2 \gamma)}
\left(3+\left(1+ \frac{n_f}{N_c^3}\right)\frac{(2+3\gamma(1-\gamma))}{(3-2\gamma)(1+2\gamma)}\right) 
+ \frac{3}{2} \zeta_3 - \frac{\beta_0}{8 N_c} \chi_0^2 \left(\gamma\right).  \nonumber
\end{eqnarray}
All these expressions are re--written here just to set the notation. 
$\psi \left(\gamma \right) = \Gamma'\left(\gamma\right)/ 
\Gamma \left(\gamma\right)$ and
\begin{eqnarray}
\phi \left(\gamma\right) + \phi \left(1-\gamma\right) &=&\nonumber\\ 
&&\hspace{-2cm}\sum_{m=0}^{\infty} \left(\frac{1}{\gamma+m}+\frac{1}{1-\gamma+m}\right)
\left(\psi'\left(\frac{2+m}{2}\right)-\psi'\left(\frac{1+m}{2}\right)\right).
\end{eqnarray}
For future reference it is important to remark that the terms with second derivatives
 in the $\psi$ function directly stem from the double logarithm in 
Eq.~(\ref{ktKernel}). The pole structure of this kernel around $\gamma = 0,1$ 
is as follows:
\begin{eqnarray}
\chi_0 \left(\gamma\right) &\simeq& \frac{1}{\gamma} +
\left\{\gamma \rightarrow 1- \gamma\right\},\\
\chi_1 \left(\gamma \right) &\simeq& 
\frac{\rm a}{\gamma}+\frac{\rm b}{\gamma^2}-\frac{1}{2 \gamma^3} +
\left\{\gamma \rightarrow 1- \gamma\right\}.
\end{eqnarray}
The cubic poles come directly from $\psi''$ and 
\begin{eqnarray}
\label{ab}
{\rm a} &=& \frac{5}{12}\frac{\beta_0}{N_c} -\frac{13}{36}\frac{n_f}{N_c^3}
-\frac{55}{36}, \, \, \,
{\rm b} ~=~ -\frac{1}{8}\frac{\beta_0}{N_c} -\frac{n_f}{6 N_c^3}
-\frac{11}{12}.
\end{eqnarray}
Throughout this paper for numerical analysis the value $n_f = 3$ will be taken. 

In this $\gamma$--space representation it is known that the cubic poles do 
compensate for the equivalent terms appearing when the symmetric 
Regge--like energy scale $s_0 = q_1 q_2$ is shifted to the $s_0 = q_{1,2}^2$ 
choice in DIS--like processes. This compensation takes place at 
NLL, the accuracy at which the BFKL kernel is known. It turns out that higher 
order terms beyond NLL, not compatible with RG evolution, 
are also generated by this change of scale. The NLL truncation of the 
perturbative expansion is then the reason why the gluon Green's function 
in the BFKL formalism develops oscillations in the $q_1^2/q_2^2$ ratio when 
this ratio is very far from unity. Along these oscillations the Green's 
function can have negative values.

At present collider energies it remains to be seen how important 
these oscillations are when a full NLL BFKL--resummed calculation of a 
physical cross section, including impact factors, 
is carried out. Nevertheless there have been attempts in the literature to 
improve the behaviour of the BFKL resummation in the $q_1^2/q_2^2$ variable. 
One of the original proposals, the subject of the present analysis, shows 
how it is possible to remove the most dominant 
poles in $\gamma$--space incompatible with RG evolution by simply shifting 
the $\omega$--pole present in the BFKL scale invariant 
eigenfunction. This shift has to be 
performed with care not to double count terms and to match the NLL accuracy 
of the original BFKL calculation. In the present study the focus will be 
on the scheme proposed in Ref.~\cite{Salam}:
\begin{eqnarray}
\omega &=& \nonumber\\
&&\hspace{-1.1cm}{\bar \alpha}_s \left(1+\left({\rm a}+\frac{\pi^2}{6}\right){\bar \alpha}_s\right) \left(2 \psi(1)-\psi\left(\gamma+\frac{\omega}{2}-{\rm b}\,{\bar \alpha}_s \right)-\psi\left(1-\gamma+\frac{\omega}{2}-{\rm b}\,{\bar \alpha}_s \right)\right)\nonumber\\
&&\hspace{-1cm}+ {\bar \alpha}_s^2 \left(\chi_1 \left(\gamma\right) 
+\left(\frac{1}{2}\chi_0\left(\gamma\right)-{\rm b}\right)\left(\psi'(\gamma)+\psi'(1-\gamma)\right)-\left({\rm a}+\frac{\pi^2}{6}\right)\chi_0(\gamma)\right).
\end{eqnarray}
By iterating this expression new terms beyond the original BFKL calculation 
appear improving the convergence in $q_1^2/q_2^2$ of the expansion. It is 
worth noting that the shift immediately resums these new terms to all orders 
in the coupling. Therefore, any attempt to modify this approach should 
retain this feature.

\section{``All--poles'' resummation}
\label{AllPoles}

The main idea underlying the present analysis is that the numerical solution 
to the equation
\begin{eqnarray}
\label{simpleshift}
\omega &=& {\bar \alpha}_s \left(2 \psi(1)-\psi\left(\gamma+\frac{\omega}{2}\right)-\psi\left(1-\gamma+\frac{\omega}{2}\right)\right)
\end{eqnarray}
can be approximated remarkably well by the all--orders resummed expression
\begin{eqnarray}
\label{simplecase}
\omega &=& \int_0^1 \frac{dx}{1-x}\left\{\left(x^{\gamma-1}+x^{-\gamma}\right)
\sqrt{\frac{2 {\bar \alpha}_s}{\ln^2{x}}} J_1\left(\sqrt{2 {\bar \alpha}_s \ln^2{x}}\right)-2 {\bar \alpha}_s\right\},
\end{eqnarray}
where $J_1 (z)$ is the Bessel function of the first kind. 

\begin{figure}[tbp]
  \centering
  \epsfig{width=5cm,file=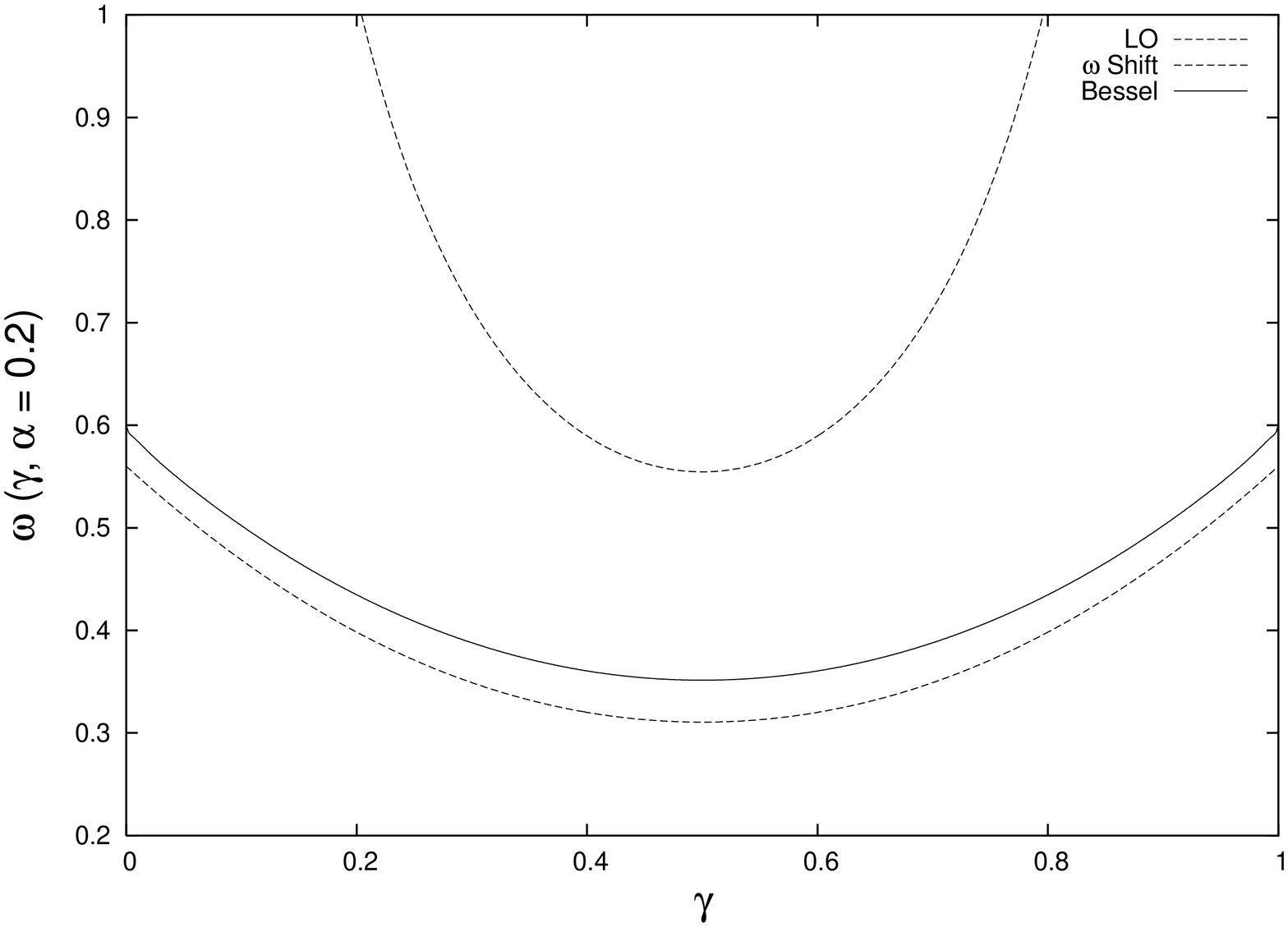,angle=-90}
  \epsfig{width=5cm,file=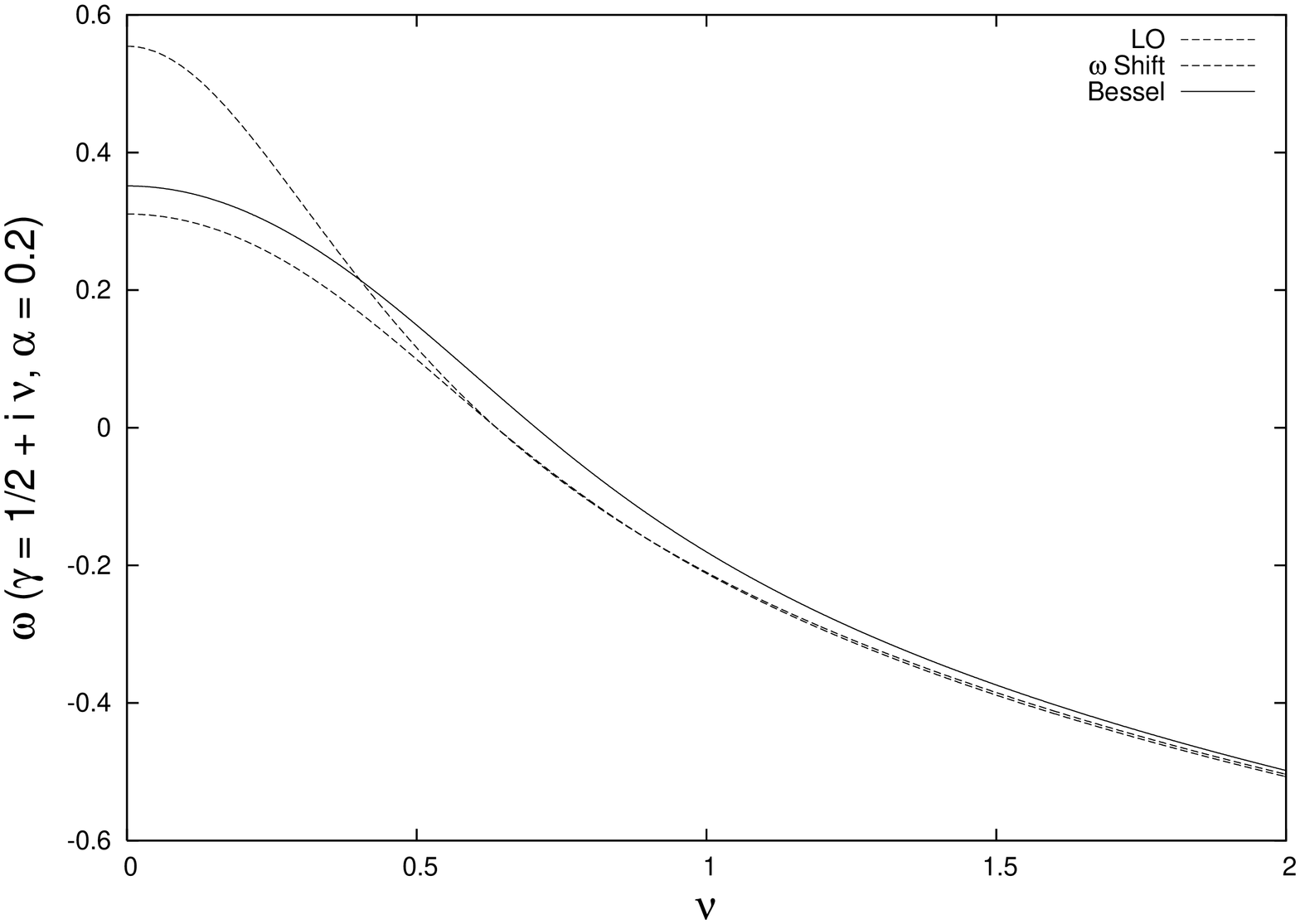,angle=-90}
  \epsfig{width=5cm,file=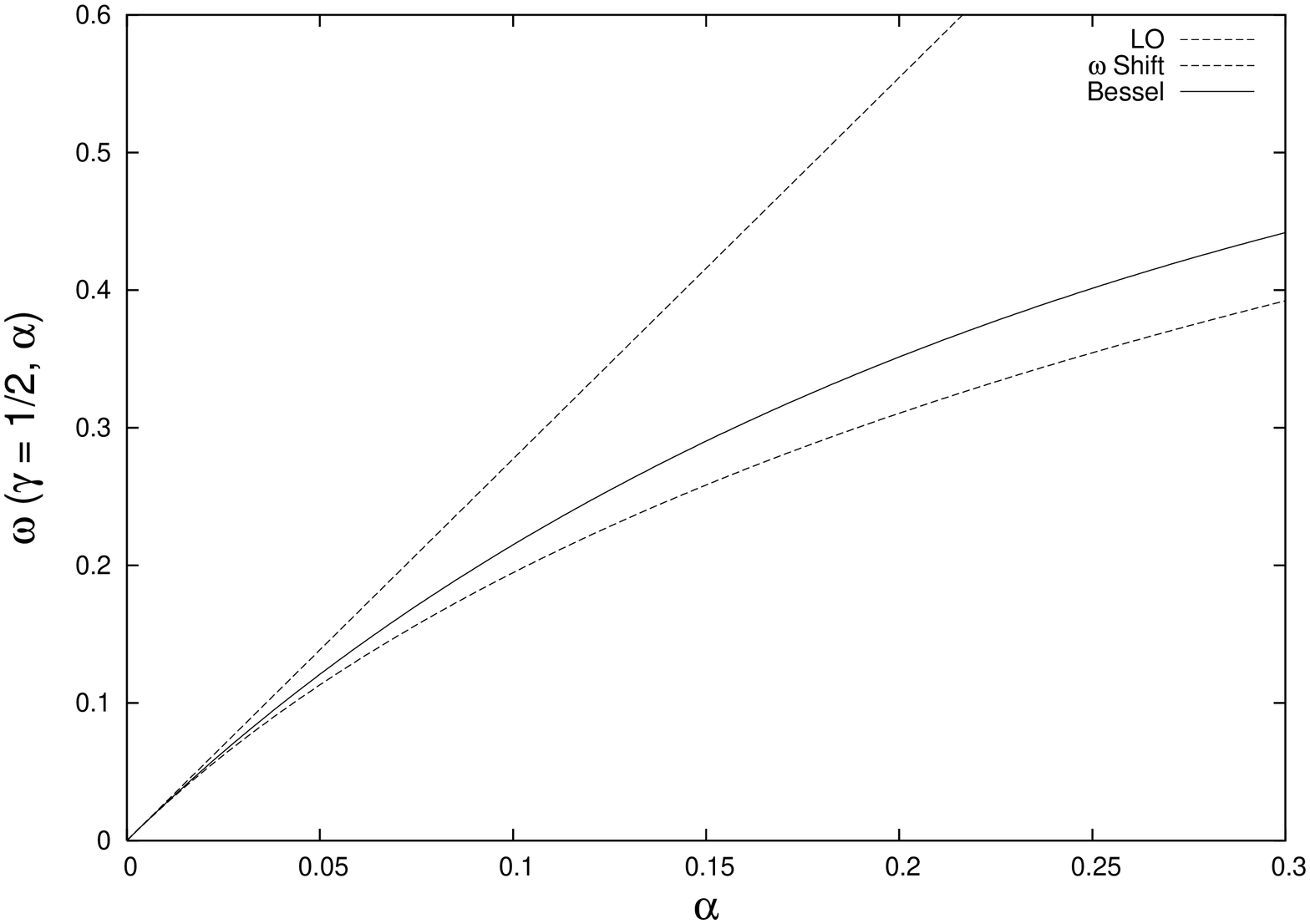,angle=-90}
  \caption{Behaviour of the $\gamma$--representation of the LL BFKL kernel compared to the $\omega$--shift of Eq.~(\ref{simpleshift}). For a fixed coupling of 
$\bar{\alpha}_s = 0.2$ on the top and middle plots around the saddle point, 
and at the dominant region $\gamma = 1/2$ for different perturbative 
values of the coupling (bottom). The approximation to the shift using the 
Bessel function resummation as in Eq.~(\ref{simplecase}) 
is also shown in the plots.}
\label{SimpleFigure}
\end{figure}

Before showing the calculations leading to this formula the 
expression in Eq.~(\ref{simplecase}) is compared to the 
numerical solution of Eq.~(\ref{simpleshift}) in Fig.~\ref{SimpleFigure}.
The values corresponding to the LL BFKL kernel are also included in the plot. 
The $\omega$--shift removes the poles at $\gamma = 0,1$ and affects the 
saddle point region by reducing the LL intercept from about 0.55 at a 
fixed coupling of $\bar{\alpha}_s = 0.2$ to a value close to 0.31. The 
approximated value obtained from the expression containing $J_1(z)$ is 0.35. 
It will be shown below how this approximation to the shift is even more 
accurate when the full NLL corrections are included. On the last plot of 
Fig.~\ref{SimpleFigure} one can see that the approximation is a stable one 
under variations of the strong coupling values.

In the following the full NLL scale invariant terms are taken into account. 
In general one can consider the $\omega$--shift of the form
\begin{eqnarray}
\omega = {\bar \alpha}_s  \left(1+ {\rm A} \,{\bar \alpha}_s \right)
\left(2 \psi(1)-\psi\left(\gamma+\frac{\omega}{2}+ {\rm B} \,{\bar \alpha}_s\right)-\psi\left(1-\gamma+\frac{\omega}{2}
+ {\rm B} \,{\bar \alpha}_s\right)\right).
\end{eqnarray}
It is convenient to use the following representation of the LL eigenvalue 
of the BFKL kernel in Eq.~(\ref{LOeigen}):
\begin{eqnarray}
\chi_0 (\gamma) &=& \sum_{m=0}^{\infty} \left(\frac{1}{\gamma + m}+
\frac{1}{1-\gamma+m}-\frac{2}{m+1}\right).
\end{eqnarray}
This representation reveals the singularity structure of 
$\chi_0 (\gamma)$ which 
has an infinite number of poles at the $\gamma = -m, 1+m$ points along the 
real axis. In this way the general shift can be written as
\begin{eqnarray}
\label{generalshift}
\omega &=& \\
&&\hspace{-1cm}{\bar \alpha}_s  \left(1+ {\rm A} {\bar \alpha}_s\right)
\sum_{m=0}^{\infty} \left(\frac{1}{\gamma + m+\frac{\omega}{2}+ {\rm B} \,{\bar \alpha}_s}+
\frac{1}{1-\gamma+m+\frac{\omega}{2}+ {\rm B} \,{\bar \alpha}_s}-\frac{2}{m+1}\right). \nonumber
\end{eqnarray}
In those regions close to the $\gamma \simeq -m$ poles a good approximation to 
Eq.~(\ref{generalshift}) is
\begin{eqnarray}
\omega &\simeq& \frac{{\bar \alpha}_s \left(1+ {\rm A} {\bar \alpha}_s\right)}{\gamma + m+\frac{\omega}{2} + {\rm B} \,{\bar \alpha}_s},
\end{eqnarray}
with the straightforward solution
\begin{eqnarray}
\omega &=& -(\gamma + m + {\rm B} \,{\bar \alpha}_s)+
\left|\gamma + m + {\rm B} \,{\bar \alpha}_s\right|
\left(1+\frac{2{\bar \alpha}_s \left(1+ {\rm A} {\bar \alpha}_s\right)}{\left(\gamma + m + {\rm B} \,{\bar \alpha}_s\right)^2}\right)^{\frac{1}{2}}.
\end{eqnarray}
The same logic applies to the regions around the $\gamma \simeq 1 + m$ poles. In the following it will be shown how, to a very good accuracy, the solution to the shift in 
Eq.~(\ref{generalshift}) can be obtained by simply adding all the 
approximated solutions at the different poles plus a term related 
to the virtual contributions of the LL BFKL kernel, {\it i.e.}
\begin{eqnarray}
\label{assumption}
\omega &=& \sum_{m=0}^{\infty} 
\left\{-(1+ 2 m + 2 \,{\rm B} \,{\bar \alpha}_s)+
\left|\gamma + m + {\rm B} \,{\bar \alpha}_s\right|
\left(1+\frac{2{\bar \alpha}_s \left(1+ {\rm A} {\bar \alpha}_s\right)}{\left(\gamma + m + {\rm B} \,{\bar \alpha}_s\right)^2}\right)^{\frac{1}{2}}\right. \nonumber\\
&&\hspace{-1cm}\left.+
\left|1-\gamma + m + {\rm B} \,{\bar \alpha}_s\right|
\left(1+\frac{2{\bar \alpha}_s \left(1+ {\rm A} {\bar \alpha}_s\right)}{\left(1-\gamma + m + {\rm B} \,{\bar \alpha}_s\right)^2}\right)^{\frac{1}{2}}
- \frac{2 {\bar \alpha}_s \left(1+ {\rm A} {\bar \alpha}_s\right)}{m+1}\right\}.
\end{eqnarray}
From now on this expression will be indicated as the ``all--poles'' approximation. In the region of interest, $0 < \gamma < 1$, Eq.~(\ref{assumption}) 
simplifies because 
\begin{eqnarray}
\left|\gamma + m + {\rm B} \,{\bar \alpha}_s\right| &=& 
\gamma + m + {\rm B} \,{\bar \alpha}_s, \\
\left|1-\gamma + m + {\rm B} \,{\bar \alpha}_s\right| &=& 1-\gamma + m + {\rm B} \,{\bar \alpha}_s,
\end{eqnarray}
and one can also introduce the all--orders expansion of the square roots:
\begin{eqnarray}
\omega &=& \sum_{m=0}^{\infty} \left\{ - \frac{2 {\bar \alpha}_s \left(1+ {\rm A} {\bar \alpha}_s\right)}{m+1} \right. \nonumber\\
&&\hspace{.2cm} \left. + \left( \sum_{n=0}^{\infty}
\frac{(-1)^n (2n)!}{2^n n! (n+1)!}\frac{\left({\bar \alpha}_s+ {\rm A} {\bar \alpha}_s^2\right)^{n+1}}{\left(\gamma + m + {\rm B} \,{\bar \alpha}_s\right)^{2n+1}}+ \left\{\gamma \rightarrow 1-\gamma\right\} \right)\right\}.
\end{eqnarray}
At the $\gamma =0,1$ poles this expansion generates the NLL terms:
\begin{eqnarray}
\omega \simeq \frac{{\bar \alpha}_s}{\gamma} + {\bar \alpha}_s^2 \left(\frac{\rm A}{\gamma}
- \frac{\rm B}{\gamma^2}- \frac{1}{2 \gamma^3} \right)+ \left\{\gamma \rightarrow 1-\gamma\right\}. 
\end{eqnarray}
Therefore, to match the original kernel at NLL, it is needed to set 
${\rm A} = {\rm a}$ and ${\rm B} = - {\rm b}$ from Eq.~(\ref{ab}).

To demonstrate that this expansion improves the convergence of the BFKL 
calculation 
in the same way as the original $\omega$--shift, it is needed to include 
the full NLL scale invariant kernel without double counting terms, {\it i.e.}
\begin{eqnarray}
\label{All-poles}
\omega &=& \bar{\alpha}_s \chi_0 (\gamma) + \bar{\alpha}_s^2 \chi_1 (\gamma) \\
&+& \left\{\sum_{m=0}^{\infty} \left[\left(\sum_{n=0}^{\infty}
\frac{(-1)^n (2n)!}{2^n n! (n+1)!}\frac{\left({\bar \alpha}_s+ {\rm a} \,{\bar \alpha}_s^2\right)^{n+1}}{\left(\gamma + m - {\rm b} \,{\bar \alpha}_s\right)^{2n+1}}\right) \right. \right. \nonumber\\&&\left.\left.-\frac{\bar{\alpha}_s}{\gamma + m} - \bar{\alpha}_s^2 \left(\frac{\rm a}{\gamma +m} + \frac{\rm b}{(\gamma + m)^2}-\frac{1}{2(\gamma+m)^3}\right)\right]+ \left\{\gamma \rightarrow 1-\gamma\right\}\right\}. \nonumber
\end{eqnarray}
\begin{figure}[tbp]
  \centering
  \epsfig{width=5cm,file=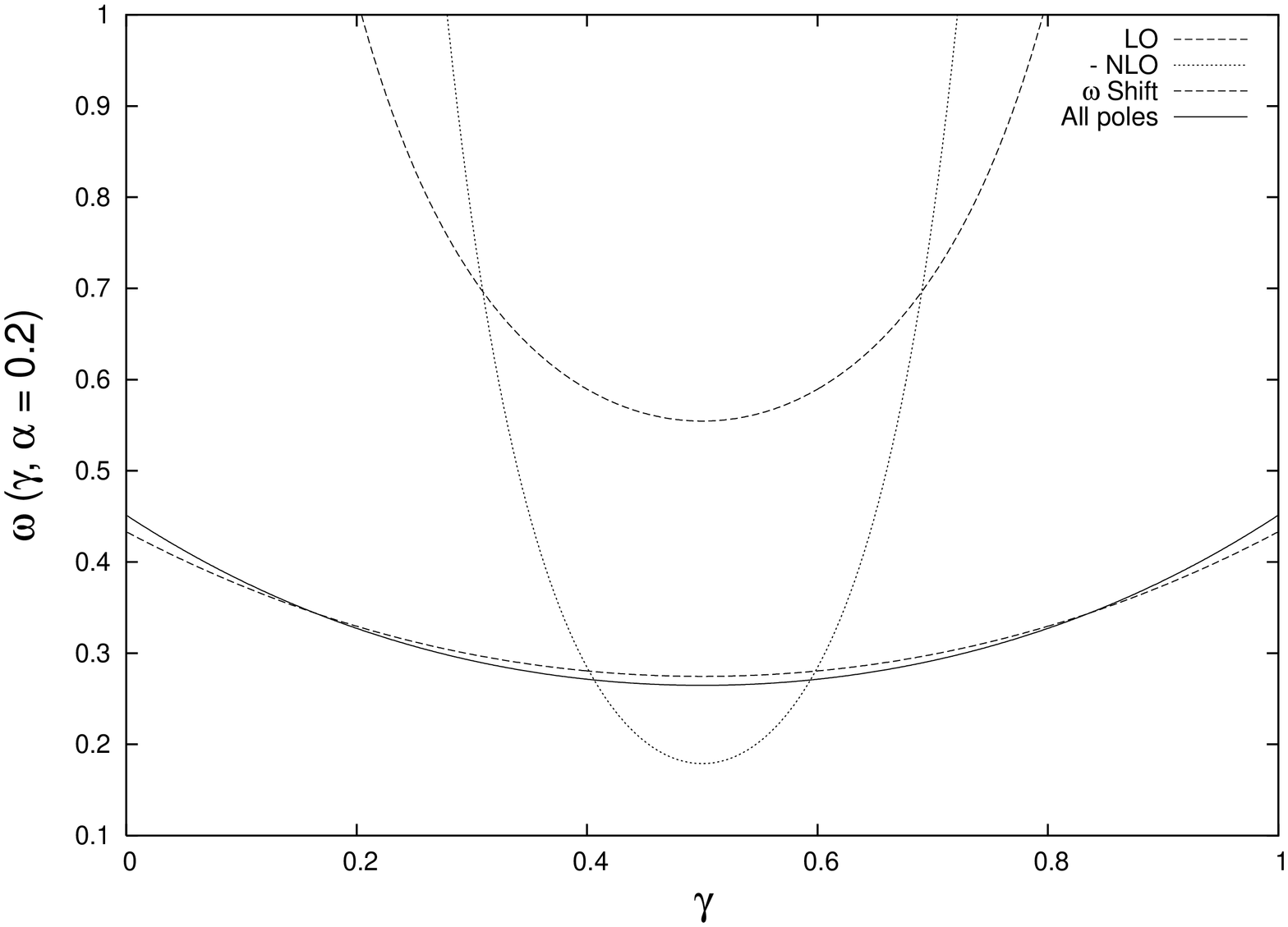,angle=-90}
  \epsfig{width=5cm,file=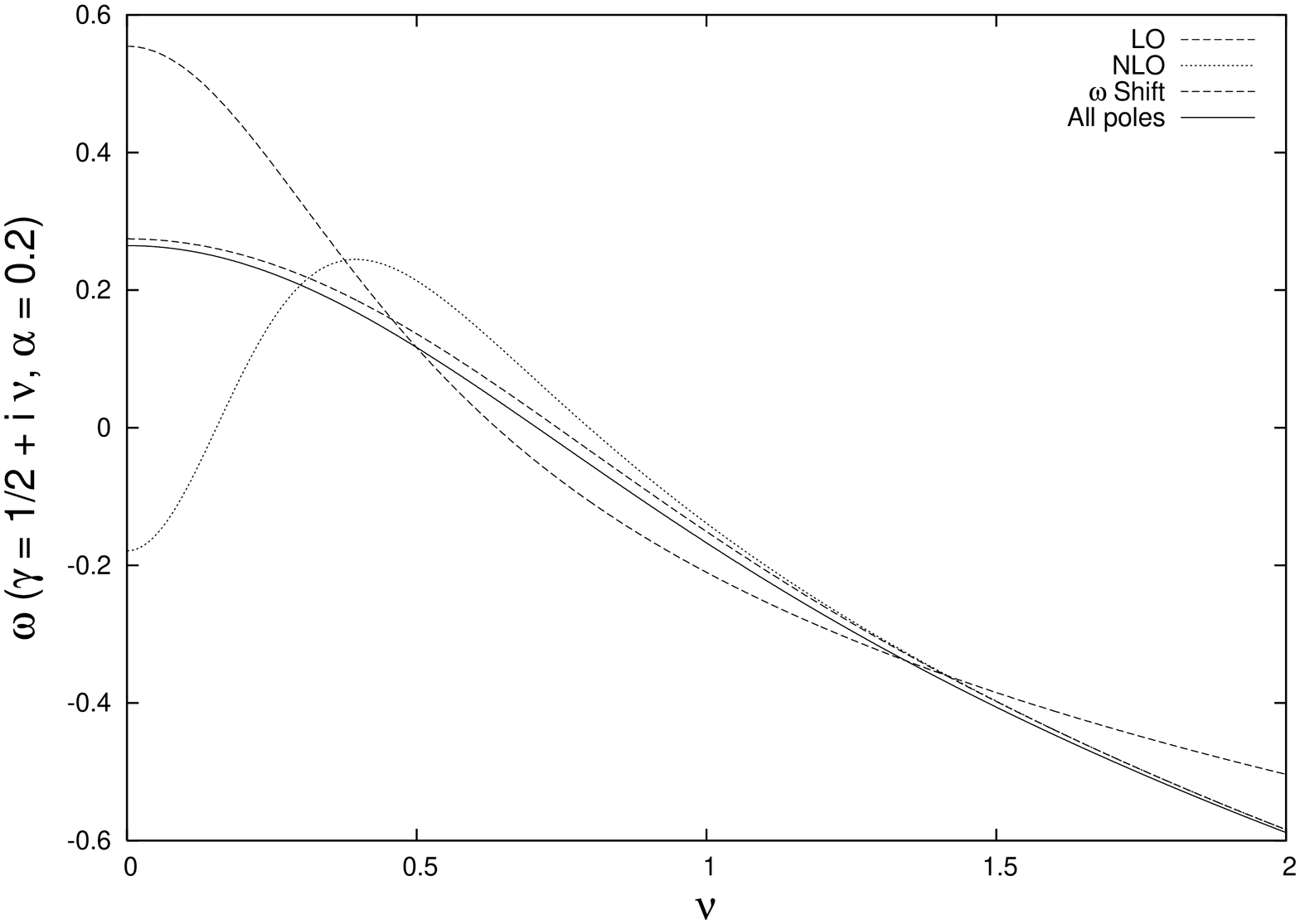,angle=-90}
  \epsfig{width=5cm,file=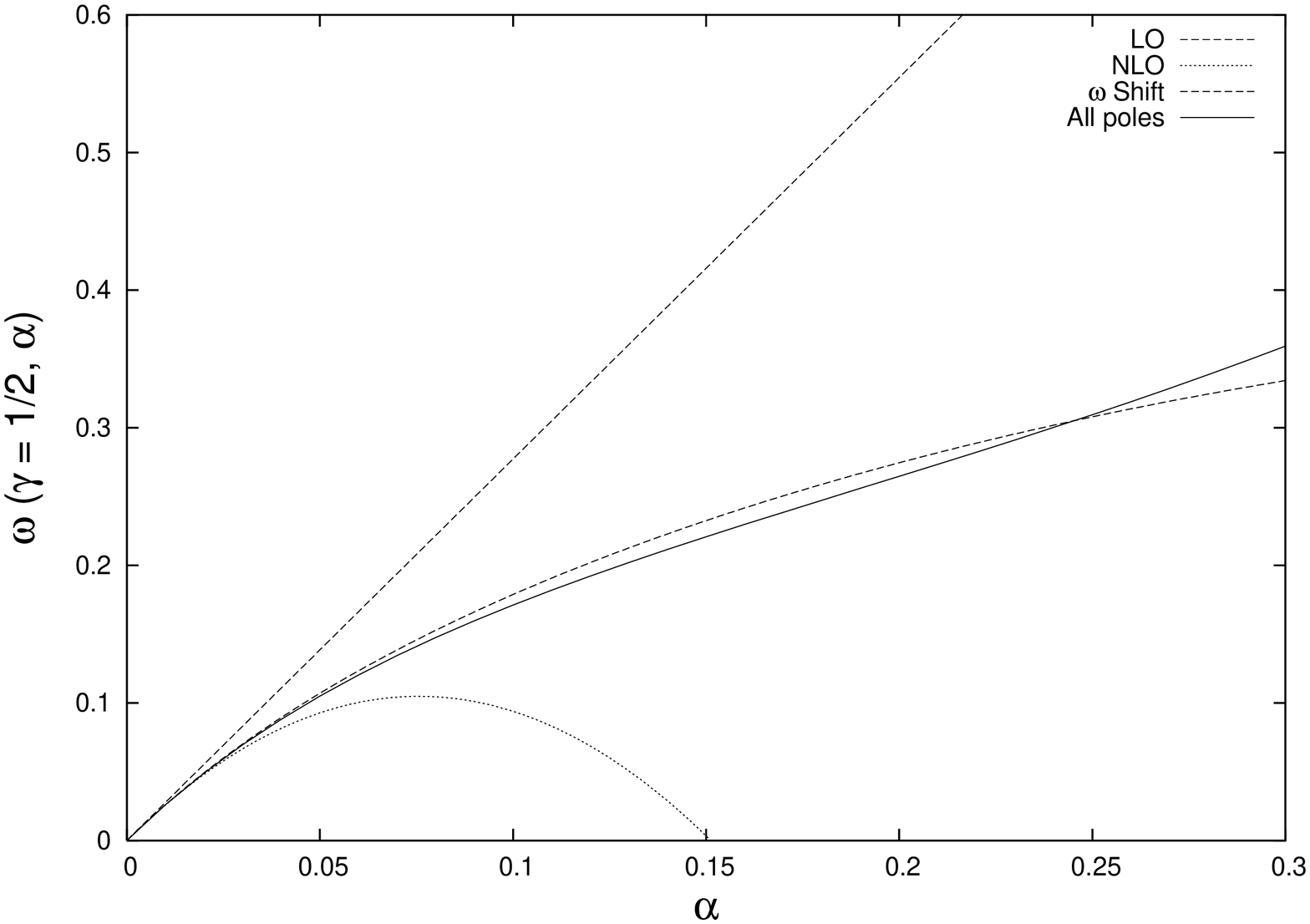,angle=-90}
  \caption{The $\gamma$--representation of the LL and NLL scale 
invariant kernels showing the instable behaviour of the last one. The 
RG--improved kernel by a $\omega$--shift is also included, 
together with the ``all--poles'' approximation proposed in the text.}
\label{FullFigure1}
\end{figure}

In this way, as in the original $\omega$--shift, the NLL results are untouched 
and only higher order terms are added to the kernel. The extra terms being 
resummed to all orders in $\bar{\alpha}_s$ will be important to find 
a closed form for the kernel with corresponding eigenvalues as in 
Eq.~(\ref{All-poles}). 

It turns out that when the matching to NLL is performed 
the ``all--poles'' result reproduces the $\omega$--shift very accurately. 
This can be seen in Fig.~\ref{FullFigure1} where the full $\omega$--shift and 
the ``all--poles'' kernel are compared, together with the LL and NLL results 
which are also included. The fact that the imaginary part of $\gamma$ at the 
maximum of the NLL scale invariant eigenvalue 
(middle plot of Fig.~\ref{FullFigure1}) is not 
zero results in the oscillations in the $q_1^2/q_2^2$ variable. This feature 
is removed when the RG--improved kernel is used, as it 
also happens for the ``all--poles'' kernel. The estimated intercept 
from the $\omega$--shift is of about 0.27, while that coming from the 
``all--poles'' kernel is 0.26 (for $\bar{\alpha}_s = 0.2$). The similarity 
between the ``all--poles'' kernel and the original $\omega$--shift remains for 
different values of the coupling (see bottom plot of Fig.~\ref{FullFigure1}).

\section{Bessel representation in transverse momentum space}
\label{AporBessel}

The most striking feature of Eq.~(\ref{All-poles}) is that the 
$\omega$--space is decoupled from the $\gamma$--representation, {\it i.e.} they no longer 
mix as it occurs in the original $\omega$--shift. This has the implication 
that it should be possible to find an expression for a RG--improved BFKL kernel 
which does not mix longitudinal with transverse 
degrees of freedom. To find such a kernel it is convenient to define 
the following function related to a simple representation of the LL kernel:
\begin{eqnarray}
\Omega(\gamma) &\equiv& 
\int_0^{\infty} \frac{d q^2}{\left|k^2 - q^2\right|}
\left(\left(\frac{q^2}{k^2}\right)^{\gamma -1 - {\rm b}{\bar \alpha}_s \frac{\left|k - q\right|}{k-q}}-2 \frac{{\rm min} \left( k^2, q^2\right)}{k^2 + q^2}\right) \\
&=& \int_0^1 \frac{d x}{1-x} \left(x^{\gamma-1-{\bar \alpha}_s {\rm b}}+x^{-\gamma-{\bar \alpha}_s {\rm b}}-2\right) \nonumber\\
&=& 2 \psi (1) - \psi (\gamma - {\bar \alpha}_s {\rm b})
- \psi (1-\gamma - {\bar \alpha}_s {\rm b})\nonumber\\
&=& \sum_{m=0}^{\infty} 
\left(\frac{1}{\gamma + m - {\rm b} \,{\bar \alpha}_s} +
\frac{1}{1-\gamma+m- {\rm b} \,{\bar \alpha}_s}-\frac{2}{m+1}\right). \nonumber
\end{eqnarray}
Keeping in mind the coefficients of the expansion in Eq.~(\ref{All-poles}) the 
following summation can be constructed:
\begin{eqnarray}
\sum_{n=1}^{\infty}
\frac{(-1)^n \left({\bar \alpha}_s+ {\rm a}\, {\bar \alpha}_s^2\right)^{n+1}}{2^n n! \, (n+1)!} \frac{d^{2n}}{d \gamma^{2n}}\Omega (\gamma) &=& \\
&&\hspace{-6cm}\int_0^{\infty} 
\frac{d q^2}{\left| k^2 - q^2\right|}
\left(\frac{ q^2}{ k^2}\right)^{\gamma -1 - {\rm b}{\bar \alpha}_s 
\frac{\left|k- q\right|}{k-q}}
\sum_{n=1}^{\infty}
\frac{(-1)^n \left({\bar \alpha}_s+ {\rm a} \, {\bar \alpha}_s^2\right)^{n+1}}{2^n n! \, (n+1)!} 
\ln^{2n}{\left(\frac{ q^2}{ k^2}\right)} \nonumber\\
&&\hspace{-5cm}= \sum_{m=0}^{\infty}\sum_{n=1}^{\infty}
\frac{(-1)^n (2n)! }{2^n n! \, (n+1)!} 
\frac{\left({\bar \alpha}_s+ {\rm a} \, 
{\bar \alpha}_s^2\right)^{n+1}}{(\gamma + m - {\rm b} \,{\bar \alpha}_s)^{2n+1}}+\left\{\gamma \rightarrow 1-\gamma\right\}. \nonumber
\end{eqnarray}
In order to include the $n=0$ case it is needed to add a term 
corresponding to the virtual contributions in the LL BFKL kernel, ensuring, 
in this way, the infrared finiteness of the final result, {\it i.e.}
\begin{eqnarray}
\label{almostBessel}
\int_0^{\infty} \frac{d q^2}{\left|k^2 - q^2\right|}
\left( \left(\frac{q^2}{k^2}\right)^{\gamma -1 - {\rm b}{\bar \alpha}_s \frac{\left|k-q\right|}{k-q}}\sum_{n=0}^{\infty}
\frac{(-1)^n \left({\bar \alpha}_s + {\rm a} \, {\bar \alpha}_s^2\right)^{n+1}}{2^n n! \, (n+1)!} \ln^{2n}{\left(\frac{q^2}{k^2}\right)} \right. \nonumber\\
&&\hspace{-10.5cm}\left.-2 \left({\bar \alpha}_s+ {\rm a} \, {\bar \alpha}_s^2\right) \frac{{\rm min}(k^2,q^2)}{k^2+q^2} \right) ~=~ - \sum_{m=0}^{\infty} \frac{2 \left({\bar \alpha}_s+ {\rm a} \,{\bar \alpha}_s^2\right)}{m+1} \nonumber\\
&&\hspace{-11cm}+ \left(\sum_{m=0}^{\infty}\sum_{n=0}^{\infty}
\frac{(-1)^n (2n)! }{2^n n! \, (n+1)!} 
\frac{\left({\bar \alpha}_s+ {\rm a} \, {\bar \alpha}_s^2\right)^{n+1}}{(\gamma + m - {\rm b} \,{\bar \alpha}_s)^{2n+1}}+\left\{\gamma \rightarrow 1-\gamma \right\}\right).
\end{eqnarray}
By introducing this extra $n=0$ piece it is then possible to resum the 
expansion in terms of a Bessel function and Eq.~(\ref{almostBessel}) is 
equivalent to 
\begin{eqnarray}
\label{Besselya}
\int_0^{\infty} \frac{d q^2}{\left|k^2 - q^2\right|}
\left(-2 \left({\bar \alpha}_s+ {\rm a} \,{\bar \alpha}_s^2\right)
\frac{{\rm min}(k^2,q^2)}{k^2+q^2} \right.\\
&&\hspace{-7cm}\left.+\left(\frac{q^2}{k^2}\right)^{\gamma -1 - {\rm b}{\bar \alpha}_s 
\frac{\left|k-q\right|}{k- q}}
\sqrt{\frac{2\left({\bar \alpha}_s+ {\rm a} \,{\bar \alpha}_s^2\right)}{\ln^2{\left(\frac{q^2}{k^2}\right)}}} 
J_1
\left(\sqrt{2\left({\bar \alpha}_s+ {\rm a} \, {\bar \alpha}_s^2\right) 
\ln^{2}{\left(\frac{q^2}{k^2}\right)}}\right) \right). \nonumber
\end{eqnarray}
In order to match this expression to the NLL BFKL formulation it is sufficient 
to perturbatively expand Eq.~(\ref{Besselya}) up to order ${\bar \alpha}_s^2$:
\begin{eqnarray}
\left(\frac{q^2}{k^2}\right)^{-{\rm b}{\bar \alpha}_s 
\frac{\left|k-q\right|}{k-q}}
\sqrt{\frac{2\left({\bar \alpha}_s+ {\rm a} \,{\bar \alpha}_s^2\right)}{\ln^2{\left(\frac{q^2}{k^2}\right)}}} 
J_1 \left(\sqrt{2\left({\bar \alpha}_s+ {\rm a} \,{\bar \alpha}_s^2\right) 
\ln^{2}{\left(\frac{q^2}{k^2}\right)}}\right)\\
&&\hspace{-9cm} \simeq {\bar \alpha}_s + {\rm a} \, {\bar \alpha}_s^2
- \frac{{\bar \alpha}_s^2}{4} \ln^2{\left(\frac{q^2}{k^2}\right) 
- {\rm b}\,{\bar \alpha}_s^2 \frac{\left|k-q\right|}{k-q}}
\ln{\left(\frac{q^2}{k^2}\right)} + \dots \nonumber
\end{eqnarray}
It is important to note that this expression does not depend on the angle 
between the $\vec{q}$ and $\vec{k}$ two--dimensional vectors, therefore this 
scale invariant 
resummation can be directly added to the original NLL BFKL kernel without 
having to angular average. The ``all--poles'' approximation only affects the 
zero conformal spin sector of the theory. This is particularly interesting for 
the study of final states where angular correlations are relevant. Moreover, 
these new higher order terms do not affect the cancellation of infrared 
divergences in the original kernel.

It is also noteworthy to point out how the perturbative expansion of the 
``all--poles'' kernel naturally generates the double logarithm already 
present in the original BFKL kernel (see Eq.~(\ref{ktKernel})). What this resummation then achieves (as in the 
original $\omega$--shift) is to resum to all orders those most dominant 
transverse logarithms which will make the expansion compatible with 
renormalization group evolution in the collinear and anticollinear limits 
relevant in DIS--like configurations.

For the sake of clarity, the only modification needed in the full NLL 
kernel to introduce the ``all--poles'' resummation is to remove the term
\begin{eqnarray}
\label{presc1}
-\frac{\bar{\alpha}_s^2}{4}\frac{1}{(\vec{q}-\vec{k})^2}
\ln^2\left({\frac{q^2}{k^2}}\right)
\end{eqnarray}
 in the real emission kernel, ${\cal K}_r \left(\vec{q},\vec{k}\right)$, 
and replace it with
\begin{eqnarray}
\label{presc2}
\frac{1}{(\vec{q}-\vec{k})^2} \left\{\left(\frac{q^2}{k^2}\right)^{-{\rm b}{\bar \alpha}_s 
\frac{\left|k-q\right|}{k-q}}
\sqrt{\frac{2\left({\bar \alpha}_s+ {\rm a} \,{\bar \alpha}_s^2\right)}{\ln^2{\left(\frac{q^2}{k^2}\right)}}} 
J_1 \left(\sqrt{2\left({\bar \alpha}_s+ {\rm a} \,{\bar \alpha}_s^2\right) 
\ln^{2}{\left(\frac{q^2}{k^2}\right)}}\right) \right.\nonumber\\
&& \left.\hspace{-7cm}- {\bar \alpha}_s - {\rm a} \, {\bar \alpha}_s^2
+ {\rm b} \, {\bar \alpha}_s^2 \frac{\left|k-q\right|}{k-q}
\ln{\left(\frac{q^2}{k^2}\right)} \right\}.
\end{eqnarray}
From the asymptotic behaviour of the Bessel function, when the coupling is 
small and the difference between the $q^2$ and $k^2$ scales is not 
very large then 
\begin{eqnarray}
J_1 \left(\sqrt{2 {\bar \alpha}_s 
\ln^{2}{\left(\frac{q^2}{k^2}\right)}}\right) &\simeq& 
\sqrt{\frac{{\bar \alpha}_s}{2} \ln^{2}{\left(\frac{q^2}{k^2}\right)}},
\end{eqnarray}
and the influence of the ``all--poles'' resummation is minimal. Therefore 
the ``Regge--like'' region in momentum space is not largely affected. But 
when the logarithm in the ratio of transverse momenta becomes larger due to a 
larger difference between the scales entering the real production vertex, 
then the asymptotic behaviour develops the oscillatory form
\begin{eqnarray}
J_1 \left(\sqrt{2 {\bar \alpha}_s  
\ln^{2}{\left(\frac{q^2}{k^2}\right)}}\right) \simeq 
\left(\frac{2}{\pi^2 {\bar \alpha}_s  
\ln^{2}{\left(\frac{q^2}{k^2}\right)}}\right)^{\frac{1}{4}}
\cos{\left(\sqrt{2 {\bar \alpha}_s  
\ln^{2}{\left(\frac{q^2}{k^2}\right)}}-\frac{3\pi}{4}\right)}
\end{eqnarray}
automatically compensating for the unphysical oscillations present in the 
original formulation of the NLL BFKL equation.

\section{Change of renormalization scheme}
\label{abitmore}

\begin{figure}[tbp]
  \centering
  \epsfig{width=5cm,file=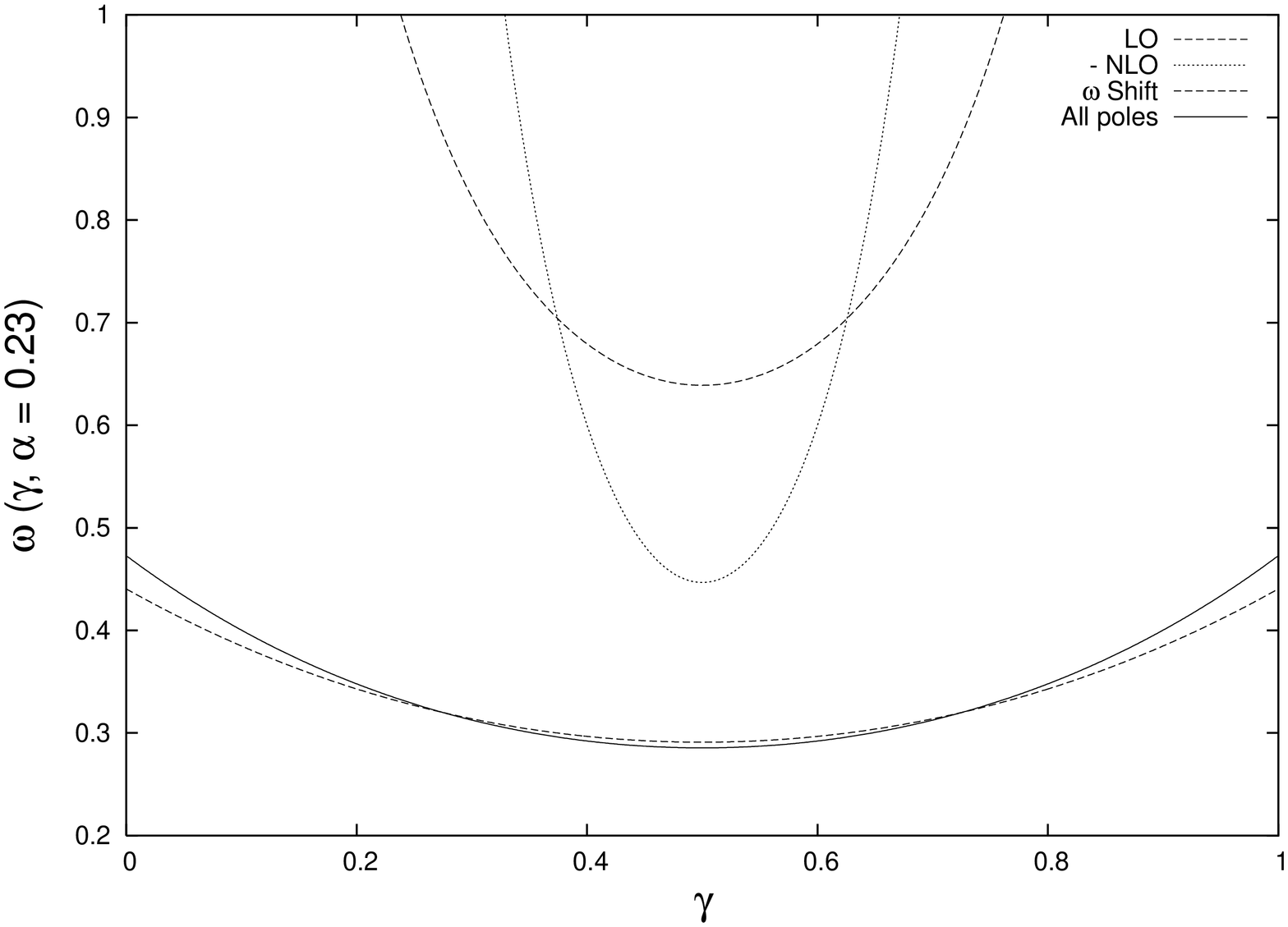,angle=-90}
  \epsfig{width=5cm,file=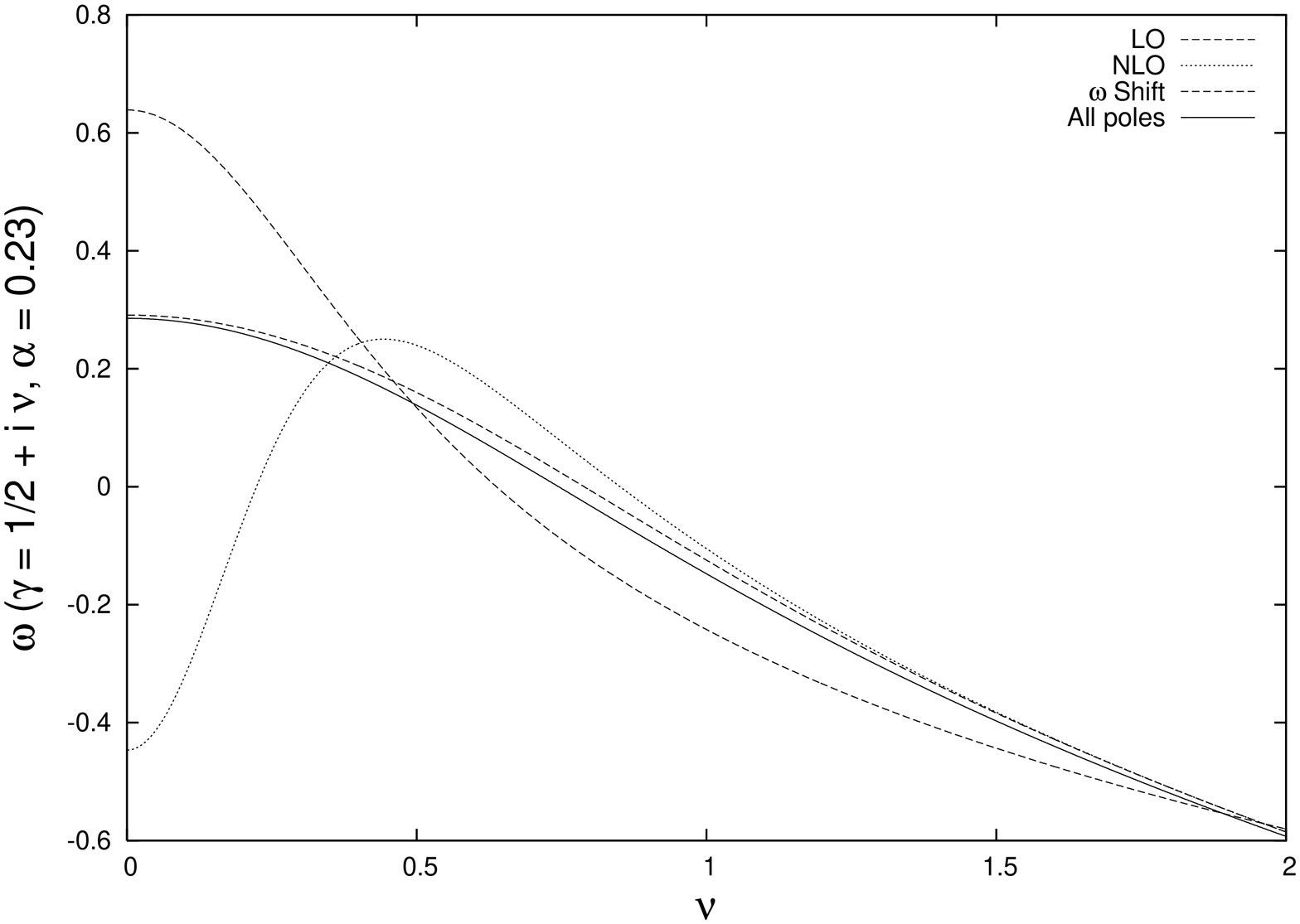,angle=-90}
  \epsfig{width=5cm,file=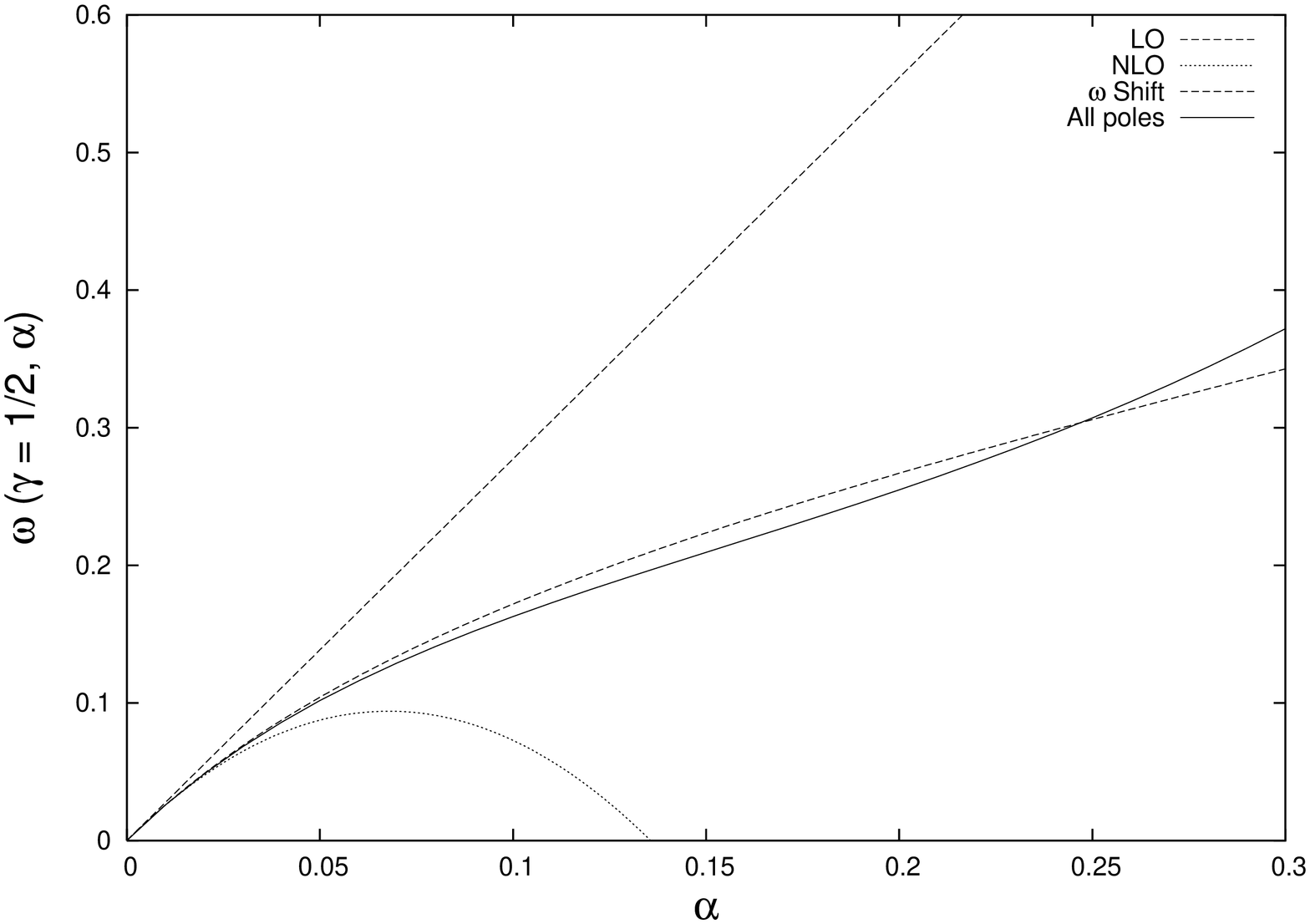,angle=-90}
  \caption{The $\gamma$--representation of the LL and NLL scale 
invariant kernels in the GB scheme. The 
RG--improved kernel using a $\omega$--shift and 
the ``all--poles'' kernel are also included.}
\label{FullFigure2}
\end{figure}

In this section the study of the effect of changing 
the renormalisation point, and using the gluon--bremsstrahlung (GB) 
instead of the 
$\overline{\rm MS}$ scheme, is performed. This is equivalent to the following 
redefinition of the position of the Landau pole: 
\begin{eqnarray}
\Lambda_{\rm GB} &=& \Lambda_{\rm \overline{\rm MS}} 
\exp{\left({\cal S}\frac{2 N_c}{\beta_0}\right)}.
\end{eqnarray}
With this choice the coefficient for the double poles in the NLL kernel does 
not change while the one for the simple poles now reads
\begin{eqnarray}
{\rm a} &=& \frac{\pi^2}{12} -\frac{13}{36}\frac{n_f}{N_c^3}
-\frac{67}{36}.
\end{eqnarray}
In Fig.~\ref{FullFigure2} this GB scheme is under analysis. There it is shown 
how again the ``all--poles'' kernel is a very good approximation to the 
$\omega$--shift. With a $\overline{\rm MS}$ value of $\bar{\alpha}_s = 0.2$ 
the corresponding coupling in GB scheme is 
$\bar{\alpha}_s^{\rm GB} \simeq 0.23$. The value of the intercept at LL is 
of $\sim 0.64$ while the RG--improved prediction is of $\sim 0.29$, which 
coincides with that steming from the ``all--poles'' analysis.
It is clear that the influence of the choice of renormalisation scheme will 
be larger when the full kernel is introduced and running coupling effects 
are also taken into account.

As a final remark, if running coupling effects are to be included, 
it is interesting to note that Eq.~(\ref{ktKernel}) corresponds to 
an angular--averaged representation of the gluon Regge trajectory and the real 
emission kernel. If this angular integration is not performed then the cancellation 
of infrared divergencies can be written as
\begin{eqnarray}
\omega_0 (q^2,\lambda) + \int \frac{d^2\vec{k}}{\pi k^2} \Gamma_{\rm cusp} 
(\bar{\alpha}_s (k^2)) \theta (k^2-\lambda^2), 
\end{eqnarray}
where the gluon Regge trajectory in this regularization reads
\begin{eqnarray}
\omega_0 (q^2,\lambda) &=& - \int^{q^2}_{\lambda^2} \frac{d k^2}{k^2}
\Gamma_{\rm cusp} (\bar{\alpha}_s (k^2)) + {\rm constant}.
\end{eqnarray}
The constant of integration can be fixed to be $\bar{\alpha}_s^2 \frac{3}{2} \zeta_3$ 
from the first reference in~\cite{FLCC} and 
\begin{eqnarray}
\Gamma_{\rm cusp} (\bar{\alpha}_s (k^2)) &=&  \bar{\alpha}_s (k^2) +
\bar{\alpha}_s^2 {\cal S},
\end{eqnarray}
is the so--called ``cusp anomalous dimension'' of Ref.~\cite{cusppaper}. 
In principle it is possible to use  
$\bar{\alpha}_s (k^2) = 4 N_c /(\beta_0 \ln{k^2/\Lambda^2})$, multiply the NLL equation 
by the extra logarithm and obtain the gluon Green's function as the solution to 
a diferential equation in 
$\gamma$--representation. An alternative is to solve the equation as proposed in Ref.~\cite{us}. 

One of the targets of the present paper is to show how it is possible to introduce a 
RG--improved kernel which still allows for an exact 
solution of the NLL BFKL equation as that proposed in Ref.~\cite{us}. The solution 
there calculated is 
based on the iteration of the BFKL equation for the partial wave expansion 
of the Green's function. This iteration has a multiple pole structure in the 
complex $\omega$--space which allows for a simple inversion back into 
energy space. This inversion using a Mellin transform would not be 
possible if an extra $\omega$--dependence in the kernel, as the one in the 
$\omega$--shift, is present. In this context the prescription proposed here 
in Eq.~(\ref{presc1}) and Eq.~(\ref{presc2}) allows for a straightforward 
implementation of the RG improvements given that it just 
modifies the real emission part of the NLL BFKL kernel without introducing 
any extra $\omega$--dependence. Therefore the iterative procedure 
used in Ref.~\cite{us} remains valid. 
In particular, if a different choice for the argument in the running coupling is considered 
the method used in  Ref.~\cite{us} always provides the 
exact solution. The study of all these effects at the level of the gluon 
Green's function will be performed in a coming publication.

\section{Conclusions}

The original analysis of Ref.~\cite{Salam} for the improvement of the convergence 
in transverse momentum space of the NLL BFKL kernel is revisited. An 
intrinsic feature of such renormalization group improved kernels is that 
the variables Mellin--conjugated of the energy and the transverse momenta, 
$\omega$ and $\gamma$ respectively, do mix due to the so--called 
$\omega$--shift. This makes the study of these schemes quite complicated. 
In the present work it is shown how it is possible to avoid the difficulties 
of the $\omega$--shift without loosing the physical insight. This is done 
by approximating the solution to the original equation at each of the 
poles of the original kernel. The result in $\gamma$--representation
 is that of 
Eq.~(\ref{All-poles}) named as the ``all--poles'' resummation. It turns out 
that the corresponding scale invariant kernel in transverse momentum space 
with eigenvalue as in  Eq.~(\ref{All-poles}) can be found and has a very 
simple form. Its structure in terms of a Bessel function of the first kind 
is shown in Eq.~(\ref{presc2}). The main feature of this kernel is that it 
only depends on transverse momenta and not on longitudinal degrees of freedom. 

This representation of the collinear improved kernels can be implemented 
immediately using the iterative approach described in 
Ref.~\cite{us}. The advantage of this approach is that it takes into account 
all physical effects, including the running of the coupling, exactly. It 
also allows for the study of final states and angular correlations.

\begin{flushleft}
{\bf \large Acknowledgements}
\end{flushleft}
The author would like to acknowledge Jeppe Andersen, Victor Fadin, Gregory 
Korchemsky, Gavin Salam, Raju Venugopalan and, in particular, Jochen Bartels, Leszek Motyka and Lev Lipatov, for discussions.

\end{document}